%% file: main.tex
\documentclass[aps,prx,twocolumn,superscriptaddress,10pt]{revtex4-1}
 
\usepackage{physics}
\usepackage{overpic}
\usepackage[caption=false]{subfig}
\usepackage[x11names]{xcolor}
\usepackage{amssymb}
\usepackage{amsthm}
\usepackage{paralist}

\newcommand{\su}[1]{\text{SU}(#1)}
\newcommand{\bigO}{\mathcal{O}}

\newcommand{\eset}{\mathbf{C}}
\newcommand{\tset}{\mathbf{T}}
\newcommand{\noise}{\mathcal{E}}
\newcommand{\avg}{\mathbb{E}}

\newtheorem{theorem}{Theorem}

\begin{document}

\title{Concepts and conditions for error suppression through randomized compiling}

\author{Adam Winick}
\affiliation{Quantum Benchmark Inc., 51 Breithaupt Street\\ Suite 100, Kitchener, ON N2H 4C3, Canada}
\affiliation{Institute for Quantum Computing, 200 University Ave W\\ Waterloo, ON N2L 3G1, Canada}
\author{Joel J. Wallman}
\affiliation{Quantum Benchmark Inc., 51 Breithaupt Street\\ Suite 100, Kitchener, ON N2H 4C3, Canada}
\affiliation{Institute for Quantum Computing, 200 University Ave W\\ Waterloo, ON N2L 3G1, Canada}
\author{Dar Dahlen}
\author{Ian Hincks}
\author{Egor Ospadov}
\affiliation{Quantum Benchmark Inc., 51 Breithaupt Street\\ Suite 100, Kitchener, ON N2H 4C3, Canada}
\author{Joseph Emerson}
\affiliation{Quantum Benchmark Inc., 51 Breithaupt Street\\ Suite 100, Kitchener, ON N2H 4C3, Canada}
\affiliation{Institute for Quantum Computing, 200 University Ave W\\ Waterloo, ON N2L 3G1, Canada}
\affiliation{Canadian Institute for Advanced Research,
661 University Ave\\Toronto, ON M5G 1M1, Canada}

\begin{abstract}
Randomized compiling reduces the effects of errors on quantum computers by tailoring arbitrary Markovian errors into stochastic Pauli noise. Here we prove that randomized compiling also tailors non-Markovian errors into local stochastic Pauli noise and investigate the technique's limitations. We show through analysis and numerical results that randomized compiling alters errors in three distinct helpful ways. First, it prevents the coherent accumulation of errors (including hard to remove crosstalk effects) across gate cycles by destroying intercycle coherent correlations. Second, it converts individual gate cycle errors into Pauli noise. Finally, randomized compiling reduces the variability inherent to noisy devices. We confirm these theoretical predictions with the IBM Quantum Experience platform and describe experimental data that illustrates a drastic performance improvement across public devices. These results cement the importance of randomized compiling in near- and long-term quantum information processing.
\end{abstract}

\maketitle

\section{Introduction}

Quantum computers exploit distinctly quantum properties of systems to solve problems much faster than digital counterparts \cite{Grover1996,Shor1997}. However, engineering large-scale devices that process quantum information has proven exceptionally difficult. Decoherence and imperfect control limit the coherent manipulation of large ensembles of particles. While quantum error correction (QEC) \cite{Shor1995,Gottesman1996,Knill1998} provides robust schemes for executing quantum algorithms on error-prone systems, the methods usually assume that the errors are well-behaved and lies below some threshold \cite{Aharonov1999,Steane2003}. The burden of QEC can be substantial, and reaching error rates well below these thresholds can dramatically improve the processing capabilities of a device.

The sequence of gates that implements a quantum algorithm is not unique, and the error rate associated with each sequence can vary drastically. Due to the massive number of gate sequences and the inherent complexity of real error processes, finding an optimal circuit is all but impossible. Alas, even estimating the worst-case error rate \cite{Kitaev2003} for large systems is intractable. Moreover, the definition of an error rate is ambiguous and depends on the figure of merit, which itself depends on the application. Even typical \cite{Wallman2015a} error rates are hard to estimate. There exists a hierarchy of error processes with increasingly desirable properties at the cost of realism and generality. For example, quantum circuits subject to Markovian errors typically have higher error thresholds than the same circuits under general errors. Stochastic Pauli noise is a subset of Markovian errors that is both efficiently characterizable \cite{Emerson2005,Emerson2007b,Knill2008,Dankert2009,Erhard2019,Flammia2019,Harper2019} and affords considerably higher fault-tolerance thresholds than Markovian errors \cite{Aharonov1999,Aliferis2007,Knill2005,Wang2011}. In addition to circuit optimization, we might also modify properties of the errors using randomized quantum circuits with varying constraints and effectiveness \cite{Knill2004,Kern2005,Wallman2016a}. Randomized compiling (RC), which is the only known method for transforming generic errors affecting \emph{universal} quantum circuits, was shown to tailor Markovian errors affecting individual gate cycles (e.g., a set of simultaneous gates) with arbitrary coherence and spatial correlations into stochastic Pauli noise. Remarkably, the technique works \emph{without} requiring additional circuit depth, although there is a small amount of digital precompilation. We note that stochastic noise and the effective stochastic noise channel induced by RC may be further corrected \cite{Li2017}.

In this paper, we investigate the efficacy of RC on systems with time-dependent and non-Markovian errors. We identify necessary conditions for the accumulation of highly coherent non-Markovian errors and argue that RC is useful in all but the most adversarial scenarios. We also clarify the effect of RC and how it employs the same error suppression mechanism as random dynamical decoupling (RDD) \cite{Viola2005a,Kern2005a}. There are two complementary ways through which RC helps overcome the errors affecting a quantum computer. These effects arise in the single- and many-randomized compilation limits.
\begin{enumerate}
\item single-randomization -- reduces the probability that coherent (and non-Markovian) errors compose in an adversarial fashion. The effect is comparable to RDD.
\item many-randomizations -- tailors errors into stochastic Pauli noise. Norm-based error metrics improve quadratically.
\end{enumerate}
Randomized benchmarking (RB) and its variants are the only known scalable error characterization methods and employ these same effects. Therefore, RC is not only helpful but an essential tool if RB estimates are to be applied confidently to a partially characterized system. With its negligible compilation cost, we expect RC to be an invaluable tool in the quest towards fault-tolerant quantum computation.

The paper is structured as follows. In Sec.~\ref{sec:Background}, we review the formalism for RC. In Sec.~\ref{sec:Results}, we present our results on RC under time-dependent and non-Markovian errors. In Sec.~\ref{sec:Discussion}, we explain the connection between RC and RDD and discuss the multifaceted benefits of the former. In Sec.~\ref{sec:Experiments}, we present experimental data obtained on publicly available IBM Quantum Experience devices validating the real performance gains of RC. Finally, we conclude in Sec.~\ref{sec:Conclusion}.

\section{Background} \label{sec:Background}

To implement a quantum algorithm on an experimental device, we need to compile a unitary matrix into a series of local logical operations. Unlike in modern digital computers, problematic errors seems intrinsic to quantum gates. We make a distinction between `easy' and `hard' gates to formulate a canonical circuit representation. Easy gates have a small amount of errors, while hard gates have much more. Contemporary multiqubit gates have lower fidelity than single-qubit gates, and this distinction establishes a logical separation. In the future, we might implement fault-tolerant operations in ways that motivate other divisions, e.g., transversal Paulis vs. magic state injection \cite{Bravyi2005} or code deformations \cite{Bombin2009}. With the division, we can reorganize any circuit into $M$ clock cycles, where each cycle consists of a round of easy and hard gates, and we assume that the circuit begins and ends with easy gates. Concretely, the $k$-th round of noisy gates reads $R_k = G_k \noise(G_k) \noise_e C_k$, where $G_k$ and $C_k$ denote hard and easy gate rounds respectively. The error process $\noise(G_k)$ is the hard gate errors and may depend on $G_k$. The process $\noise_e$ is the easy gate errors, and we assume for the moment that it does not depend explicitly on the choice of $C_k$.

The idea of RC is to replace each round of easy gates with random dressed gates. The $k$-th round of easy gates becomes $\tilde{C}_k = T_k C_k T_{k-1}^c$, where $T_k$ is chosen uniformly at random from a twirling set $\tset$, and $T_k^c = G_k T_k^\dagger G_k^\dagger$ undoes the randomization from the previous round. For a generic twirling set and generic hard gates, $T_k^c$ might not belong to the group $\eset$ generated by the easy gate set. The important point is that there are \emph{practical} divisions so that $T_k^c \in \eset$ for all gates $T_k$. For example, in our simulations and experiments, we set $\eset=\text{SU}(2)^{\otimes N}$, $\tset=\mathbf{P}_2^{\otimes N}$ ($\mathbf{P}_2$ is the set of 2x2 Pauli matrices) and the hard gates to be the controlled-not gate $\Delta(X) = \ketbra{0} \otimes I + \ketbra{1} \otimes X$. The gate set is both universal and native to several popular interfaces \cite{McKay2016}. In order to tailor the last gate cycle, we implement the final round of twirling gates through classical post-processing (see Ref.~\cite{Wallman2016a} for more details).

A single circuit randomization is often sufficient to prevent the buildup of highly coherent errors. Nevertheless, the error tailoring does not occur in a particular randomized circuit; the expected errors over many circuits tends towards a stochastic Pauli channel. Ref.~\cite{Wallman2016a} proved the following:
\begin{theorem} \label{theorem:RCMarkov}
Randomly sampling the twirling gates independently in each round tailors the errors affecting each cycle into stochastic Pauli noise when the errors on the easy gates is gate-independent.
\end{theorem}

The theorem establishes that the technique is robust to gate-dependent errors on the hard gates, which is the dominant form of gate-dependence. However, it requires that the easy gate errors are effectively gate-independent. In practice, there will be residual control errors that generate small gate-dependent coherent errors on the easy gates. Ref.~\cite{Wallman2016a} proved that easy gate-dependent errors introduces a relatively small additional error. The diamond distance between the noisy gate-dependent circuit and the equivalent gate-independent circuit grows linearly in circuit depth and is especially small when the twirling group $\tset$ is normalized by $\eset$.

\section{Results} \label{sec:Results}

Having reviewed RC under time-independent Markovian errors, we study its effectiveness at mitigating more general errors.

\subsection{Time-dependent errors}

The first type of errors we look at is time-dependent Markovian errrors, which researchers sometimes refer to as non-Markovian. Such processes vary with time and do not typically generate a dynamical semigroup. Nevertheless, if we describe the behavior with a two-parameter family of dynamical maps $\Phi(t_2, t_1)$, we get an analogous semigroup property
\begin{equation}
	\Phi(t+\tau,0)=\Phi(t+\tau,\tau)\Phi(\tau,0) \,.
\end{equation}
Even more generally, we consider a collection of dynamical maps $\{\Phi_a\}$, where each circuit realization samples errors from the collection. Examples in this class of processes are colored noise and time-dependent drift. The following theorem describes the effects of RC on the class of errors.

\begin{theorem} \label{theorem:RCMarkovTime}
Randomly sampling the twirling gates independently in each round tailors the time-dependent Markovian errors affecting each cycle into time-dependent stochastic Pauli noise when the errors on the easy gates are gate-independent.
\end{theorem}
\begin{proof}
Since each round of the compiled circuit is randomized independently, we can study a particular round $k$ and the associated error distribution $\{\noise_a\vert\noise_a:=\noise^{a,k}(G_k)\noise_e^{a,k}\}$. The superscript $a$ and $k$ denote the circuit realization and the gate round, respectively. Averaging over the error distribution and the twirling set are independent. Thus we can exchange the order of averaging, and the result follows from Theorem~\ref{theorem:RCMarkov}.
\end{proof}

Although the time-dependent errors affecting one gate cycle are indistinguishable from its time-independent counterpart, there may be differences in the cumulative error over several gates. If the characteristic correlation timescale $\tau$ is much smaller than the time $t_c$ for a cycle, there will be no discernable difference. In contrast, if $t_c \lesssim \tau$, consecutive Pauli weights can be temporally-correlated and exhibit behavior that will manifest in a similar way to how temporal correlations appear under RB \cite{Epstein2014,Ball2016,Fong2017}.

\subsection{Non-Markovian errors}

The next error model we examine is non-Markovian errors. We describe these type of errors by introducing a persistent ancillary subsystem. The error process $\noise(G_k)\noise_e$ now acts on the computational and ancillary subsystems, thereby modeling arbitrary non-Markovian behavior. The ancillary system endows the errors with a history-dependent action similar to the aforementioned time-dependent error model. However, the effect of the errors on the computational subsystem no longer satisfies any semigroup-like property.

To understand the effect of RC on non-Markovian errors, we need to generalize the definition of a Pauli channel. An error process $\noise$ is a local stochastic Pauli channel if the action on the reduced state of the subsystem of interest is a stochastic Pauli channel. I.e., for an input state $\rho_{SE}$ and reduced state $\rho_S$,
\begin{equation} \label{eqn:LocalPauliChannel}
    \noise(\rho_S) = \sum_{P \in \mathbf{P}^{\otimes N}} c_P(\rho_E) P \rho_S P^\dagger \,,
\end{equation}
where the coefficients $\{c_P\}$ depend on the reduced state $\rho_E$ of $E$. 

\begin{theorem} \label{theorem:RCNonMarkov}
Randomly sampling the twirling gates independently in each round tailors the errors at each time step (except the last) into local stochastic Pauli noise when the errors on the easy gates is gate-independent.
\end{theorem}
\begin{proof}
The tailored errors in the $k$-th round are
\begin{equation}
    \tilde{\noise} = \avg_T T^\dagger \noise(G_k)\noise_e T \,.
\end{equation}
Suppose the environment is $d$-dimensional. In a block matrix representation,
\begin{equation}
    T=I_E \otimes T = \underbrace{T \oplus \dots \oplus T}_{\text{$d$ terms}} \,,
\end{equation}
and the submatrices $\tilde{\noise}_{a,b}$ of $\tilde{\noise}$ are
\begin{equation}
    \tilde{\noise}_{a,b} = \avg_T T^\dagger [\noise(G_k)\noise_e]_{a,b} T \,.
\end{equation}
When $\tset = \mathbf{P}^{\otimes N}$, each submatrix is Pauli noise. From the definition of a unitary 1-design, the average is independent of the choice of $\tset$ and produces Pauli noise for any unitary 1-design. Therefore, the errors act as a Pauli channel on the system where the coefficient of each Pauli error depends on the reduced state of the environment.
\end{proof}

Note that the same argument that we employed to prove Theorem \ref{theorem:RCMarkovTime} trivially extends to non-Markovian errors.

\subsection{Gate-dependent non-Markovian errors}

Modern single-qubit gates have small gate-dependent control errors, so the assumption that the errors are independent of the implemented easy gates is unjustified. In the Markovian setting, gate-dependent errors add a relatively small linear-in-time contribution to the total error rate. Non-Markovian dynamics are notoriously difficult to analyze, and there is no apparent generalization. In fact, a non-Markovian system can induce vanishingly small errors that compound coherently over time. We argue that the problem arises from information leakage and derive necessary conditions for the accumulation of coherent errors.

We preface our analysis with a toy model where RC fails to prevent coherent non-Markovian errors. Consider a circuit that consists of a single qubit coupled to a qubit environment and for simplicity, set $G_k=C_k=I$, and sample $T_k \in \mathbf{P}_2$. When $C_k$ acts on the system qubit, the gate-dependent errors act identically on the environment with $C_k$. A gate-independent Hamiltonian $H_\text{GI}=\alpha X\otimes X$ introduces a small coherent error, where $0 < \alpha \ll 1$ and $X,Y,Z$ denote Pauli matrices. Since $T_k \otimes T_k$ commutes with $H_\text{GI}$, RC does not destroy the coherence of the errors, and the fidelity between the noisy and noiseless state will decay quadratically from unity (assuming the initial state is not an eigenstate of $H_\text{GI}$). The failure is possible because the gate-dependent error leaks two bits, revealing the choice of twirl to the non-Markovian environment.

Gate-dependent errors only needs to leak a single bit of information to enable maximally coherent errors. Let us partition the twirls into two-element sets specifying a sort of parity, e.g., $A = \{I, X\}$ and $B = \{Y, Z\}$. Then $X$ will commute and anti-commute with $A$ and $B$, respectively. Thus by modifying the above example so that the gate-dependent error acts with any element of either $A$ or $B$ depending on the parity, the gate-independent errors will produce an identical fidelity decay.

To derive a rigorous bound, we consider implementing a quantum circuit in the presence of an adversary Eve. She exercises complete control of the environment and attempts to introduce errors that compound coherently. Before implementing the randomly compiled circuit, we share the bare circuit with her. She replies with a function that maps an ideal cycle of easy and hard gates to a noisy process that is supposed to implement the cycle, but couples arbitrarily to her environmental system $E$. Conditioned on our randomly sampled twirls, we implement the associated noisy gates, which comprise her sole interaction with our system. We make two assumptions:
\begin{enumerate}
\item Eve can only learn about the twirl on round $k$ during round $k$.
\item An adversary requires at least one bit of parity information to correlate errors coherently over time.
\end{enumerate}
When Eve knows nothing about the net twirl separating two distant rounds, the collective errors across these rounds must add incoherently. Conversely, when she knows the precise parity between two rounds, she can ensure that the errors are coherently combined.

Suppose we have an $N$ qubit circuit and $M$ rounds of gates. The collection of noisy gates that Eve generates will contain $4^NM$ elements (under Pauli twirling), which we write as $\noise_k[T_k]$, where $k$ denotes the gate round. Correlating errors on an individual qubit requires strictly less information than for several qubits since we randomly sample each qubit twirl independently. Thus we look at how much information Eve can obtain during round $k$ about qubit $n$. Given a state $\rho_{AE}$ between our system $A$ and Eve's system $E$, we define
\begin{equation}
    \rho_{k,n} = \mathbb{E}_{T_k}\frac{1}{4}\sum_{i=0}^3 \ketbra{i} \otimes \noise_k[T_k: T_{k,n}\to \sigma_i]
    (\rho_{AE}) \,,
\end{equation}
where $T_k:T_{k,n}\to \sigma_i$ signifies replacing the Pauli on qubit $n$ in $T_k$ with the Pauli $\sigma_i$. The prepended system $R$ classically encodes the label of the replacement. We bound Eve's correlation with system $R$ by the entanglement-assisted conditional entropy
\begin{equation}
    \theta_{k,n} = \inf_{AE} S(R\vert E)_{k,n} \,,
\end{equation}
where the infimum is over all states $\rho_{AE}$ and $S(\cdot\vert\cdot)$ denotes the conditional quantum entropy \cite{Nielsen2002b}. One could constrain the reduced density matrix on $A$ so that it satisfies specific properties such as being close to the ideal state.

Whenever $\theta_{k, n}$ vanishes, Eve can deduce the two-bit twirl label and predict $R$ with certainty. We are interested in a necessary condition, and the minimum information that she needs is one parity bit. We can characterize this necessary information by mapping $R$ to a one-bit parity register $R'$ and bounding the entanglement-assisted conditional entropy
\begin{equation}
    \theta'_{k,n} = \inf_{R', AE} S(R'\vert E)_{k,n} \,,
\end{equation}
where $R'$ is minimized over all parity registers.

Fano's inequality \cite{Cover2012} relates $\theta_{k,n}'$ to an upper bound $p_{k,n}$ on the probability that Eve successfully identifies the parity of the round. The following theorem bounds her information about the net parity over several rounds.

\begin{theorem} \label{theorem:parityProb}
Eve's probability of guessing the correct net parity over rounds $i$ to $j$ on qubit $n$ is bounded by $p_{i:j,n}$ satisfying
\begin{equation} \label{eqn:parityprob}
    p_{i:j,n} \leq \frac{1}{2}\prod_{k=i}^j (2p_{k,n}-1)+\frac{1}{2} \,.
\end{equation}
\end{theorem}
\begin{proof}
Consider a two-state Markov process. State 1: Eve correctly predicts the relative parity after several twirls. State 2: Eve incorrectly predicts the relative parity. Before any twirls, Eve knows the frame with certainty which we model as the initial state $x^{(0)}=(1, 0)$.

The Markov chain relating $x^{(k)}$ and $x^{(k-1)}$ is
\begin{equation}
x^{(k)} = x^{(k-1)}
\begin{bmatrix}
    p_k & 1-p_k \\
    1-p_k & p_k \\
\end{bmatrix} \,,
\end{equation}
for $1 \leq k \leq M-1$. Thus over $j-i$ rounds a bound on her probability of guessing the correct relative parity is bounded by \eqref{eqn:parityprob}.
\end{proof}

The theorem implies that when $\theta_{k,n} \leq \theta_\text{c}$ for all $k$, and some critical $\theta_\text{c}$, the mutual information between Eve and the net parity of qubit $n$ decays exponentially in time. Such a circuit implementation obeys the \emph{weak memory hypothesis}: there is a characteristic gate length $M_\text{c}$ beyond which correlated coherent errors are not possible. The hypothesis likely holds in typical experimental settings since the environment is nonadversarial, and we would not expect the internal degrees of freedom to exploit twirl information coherently. Further,  when the bare circuit has many hard-gates, the environment would need to keep track of these as well.

\subsection{Discussion} \label{sec:Discussion}

RC helps mitigate the errors affecting a quantum circuit in two complementary ways. As described in the introduction, these effects emerge in the single- and many-randomized circuit compilation(s) limit, respectively. In the single-randomized compilation limit, RC prevents errors affecting different cycles from combining coherently with high probability. The result is most relevant for highly structured circuits, and its effect is equivalent to that of RDD \cite{Viola2005a}. Therefore, we refer to it as the \emph{dynamical decoupling property} of RC. In the many-randomized circuit compilations limit, RC modifies the errors differently. The off-diagonal components of the Pauli transfer matrix (PTM) representation of the error process tend to zero, and the effective channel becomes a stochastic Pauli channel. This \emph{noise tailoring property} improves both structured and unstructured circuits. However, the apparent improvement is closely related to the nonlinearity of the figure of merit. Linear metrics, including the average gate fidelity, are unaffected by averaging over many random compilations. Meanwhile, nonlinear metrics, like the diamond distance, can decrease significantly under many-randomized circuit compilations.

\subsubsection{Dynamical decoupling property}

 \begin{figure}[t]
 \begin{center}
 \begin{overpic}[width=\columnwidth]{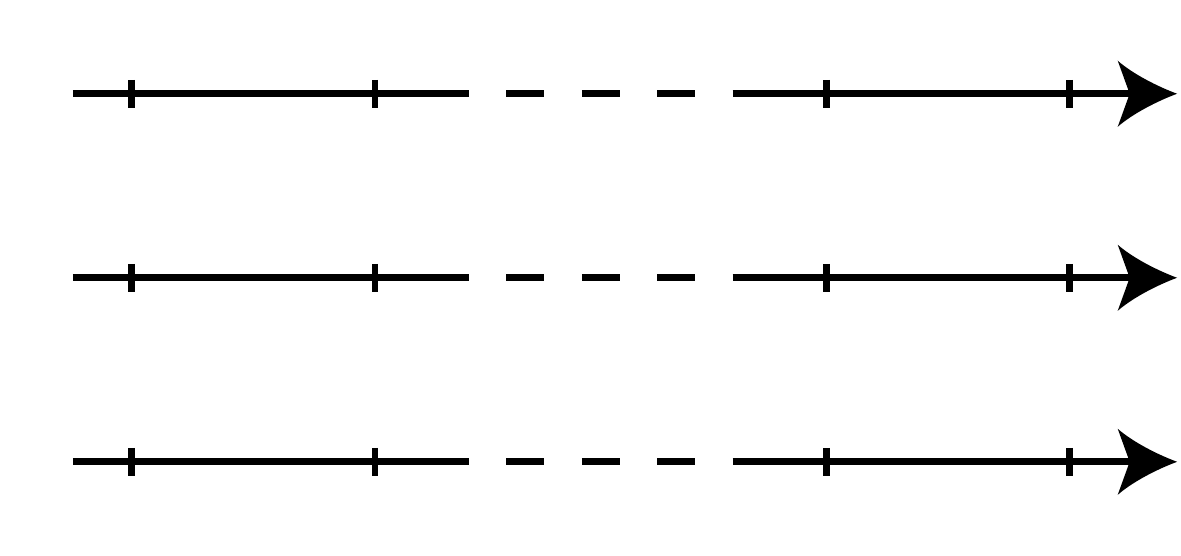}
 \put(1,38){a)}
 \put(10,42){\footnotesize $T_1$}
 \put(28,42){\footnotesize $T_1^\dagger T_2$}
 \put(64,42){\footnotesize $T_{M - 1}^\dagger T_M$}
 \put(89,42){\footnotesize $T_M^\dagger$}
 \put(10.5,34.5){\footnotesize $0$}
 \put(29,34.5){\footnotesize $\Delta t$}
 \put(61,34.5){\footnotesize $(M-1) \Delta t$}
 \put(87,34.5){\footnotesize $M \Delta t$}
 \put(1,22){b)}
 \put(7.5,26.5){\footnotesize $B_1 T_1$}
 \put(23.5,26.5){\footnotesize $T_1^\dagger B_1^\dagger B_2 T_2$}
 \put(53,26.5){\footnotesize $T_{M-1}^\dagger B_{M-1}^\dagger B_M T_M$}
 \put(85,26.5){\footnotesize $T_M^\dagger B_M^\dagger$}
 \put(10.5,19){\footnotesize $0$}
 \put(29,19){\footnotesize $\Delta t$}
 \put(61,19){\footnotesize $(M-1) \Delta t$}
 \put(87,19){\footnotesize $M \Delta t$}
 \put(1,7){c)}
 \put(7.5,11){\footnotesize $R_1 T_1$}
 \put(26,11){\footnotesize $T_1^\dagger R_2 T_2$}
 \put(60,11){\footnotesize $T_{M-1}^\dagger R_M T_M$}
 \put(89,11){\footnotesize $T_M^\dagger$}
 \put(10.5,3){\footnotesize $0$}
 \put(29.5,3){\footnotesize $\Delta t$}
 \put(61,3){\footnotesize $(M-1) \Delta t$}
 \put(87,3){\footnotesize $M \Delta t$}
 \end{overpic}
 \caption{Derivation of RC from random decoupling. a) depicts plain random decoupling. In b) each time slot incorporates an arbitrary fixed unitary within its control transformation. c) is equivalent to b) except that the last fixed unitary has been dropped so as to implement the transformation $R_M\dots R_1$.}
 \label{fig:RDEquivalence}
 \end{center}
 \end{figure}
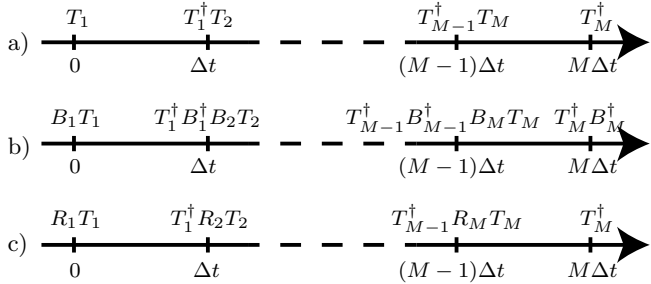

In this section, we examine the connection between RC and RDD. We also look at how RC helps reduce errors in the single randomized circuit compilation limit. Interestingly, we can understand RC as a heuristic for implementing optimized generalization of RDD \cite{Viola2005a}, which we briefly outline.

Let $S$ be a finite-dimensional system coupled to an environment $E$ that jointly evolve by the Hamiltonian $H(t)=H_0(t)+H_E$, where $H_0(t)$ denotes the part of the Hamiltonian affecting $S$, and $H_E$ is the part strictly acting on $E$. Let the available control generate a group $\mathcal{T}$ that acts noiselessly on $S$ and suppose there exists a subgroup $\tset \in \mathcal{T}$ that is a unitary 1-design. A random decoupler uses the control in a straightforward way. It selects random but known elements of $\tset$ and applies them to the system with a time $\Delta t$ between consecutive control pulses. At a time $T=M\Delta t$, we can apply the inverse of the control sequence product to recover the noisy version of the original state. Ref.~\cite{Viola2005a} proved that the average error rate $r_\text{avg}$ satisfies the theorem
\begin{theorem} \label{theorem:RDD}
Suppose that $\norm{H_0(t)}_2$ is uniformly bounded in time by $\lambda > 0$. Then RDD produces an error rate
\begin{equation}
	r_\text{avg} = \bigO\left(T\Delta t \lambda^2\right)\,,
\end{equation}
when $T\Delta t\lambda^2 \ll 1$.
\end{theorem}

Fig.~\ref{fig:RDEquivalence} schematically shows the derivation of RC from RDD. a) depicts random decoupling where $T_k$ is chosen randomly from $\tset$. In b), we make use of the fact that multiplication by a fixed unitary is a homomorphism on the set of unitary 1-designs to incorporate round-dependent unitaries $\{B_k\}$. Finally, in c) we drop $B_M^\dagger$ and retain the error suppression properties of a) and b) while realizing a net transformation $R_M\dots R_1$ where $R_k=B_kB_{k-1}^\dagger$. Effective random decoupling requires that $H_0$ is approximately independent of the choice of $T_k$ (where $H_0$ also accounts for control errors). In this way, we can understand RC as a heuristic to minimize the dependence of $H_0$ on $T_k$. With the partition into easy and hard gates, the logical hard gate lies in $R_k$, yet the random decoupling sequence suppresses its errors since it is independent of $T_k$.

The theorem for RDD is weaker than the previous results concerning RC. Pauli channels compose in accordance with Theorem \ref{theorem:RDD}, but not all such channels are Pauli channels. Nevertheless, the previous results describe the effective channel that emerges when averaging over many randomized compilations, while this result explains what happens with only a few randomized compilations.

\subsubsection{Example: Non-Markovian system}

\begin{figure}[t]
\includegraphics[width=\columnwidth]{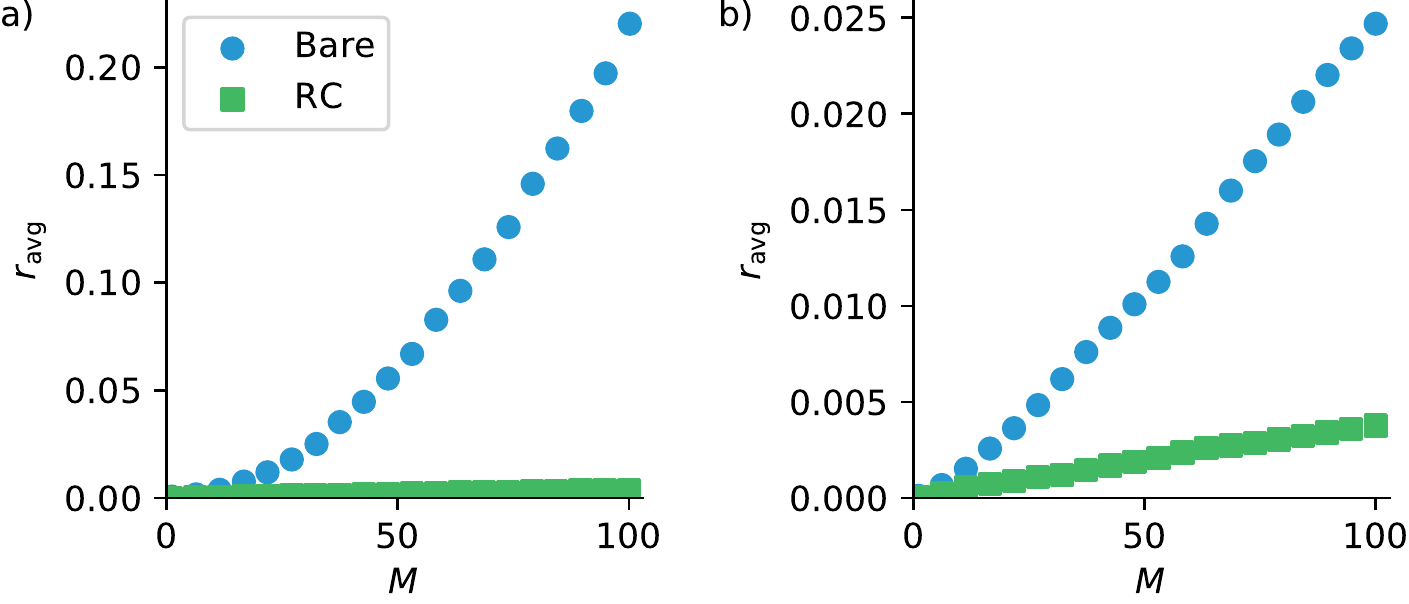}
\caption{Simulated data showing the advantage of using RC with a single-randomization to implement a highly structured circuit on six qubits that exhibit strongly non-Markovian errors. We plot the average error rate $r_\text{avg}$ as a function of the number of clock cycles $M$. With a single circuit randomization, RC (green squares) prevents the coherent non-Markovian errors from accumulating coherently and thus decreases the error rate when compared to the bare circuit (blue circles). The data points are the average of $10^2$ simulations.} \label{fig:nonmarkov}
\end{figure}

We now simulate a quantum circuit on a non-Markovian system and look at how RC's dynamical decoupling property helps overcome errors. Consider an idealized quantum dot array of $N=6$ qubits where Heisenberg interactions decay exponentially over inter-particle distance \cite{Loss1998}. The qubits do not interact directly. Instead, environmental `defect' qubits mediate inter-particle interactions, and the system-environment interactions form a 1D lattice $S-E-\dots-E-S$. The interaction Hamiltonian is
\begin{equation}
	H_I=\sum_{k=1}^{2N-1}J_k(X_kY_{k+1}+X_{k+1}Y_k) \,.
\end{equation}
We studied the dynamics of the model and found definitive signs of non-Markovianity \cite{Winick2019}.

For our numerics, we look at two different circuit configurations. Both are highly structured when compared to typical random circuits. Algorithms are inherently structured, and the example is representative of errors that might appear without error mitigation strategies. In circuit $A$, we repeatedly apply the same easy and hard round. An easy round consists of $N$ $X$ gates that act on all system qubits, and a hard round consists of $N/2$ $\Delta(X)$ gates acting on random pairs of qubits. In circuit $B$, the easy rounds are the same as in circuit $A$. The difference is that we construct hard round in $B$ by randomly sampling from the set of all hard rounds generated by the $\Delta(X)$ gate. We implement the gates in our circuit via the Hamiltonians
\begin{equation}
	H_\text{hard}=\frac{1}{10}\Delta(X), \quad\text{and}\quad H_\text{easy}=\frac{1}{2}X,
\end{equation}
that are applied for times $5\pi$ and $\pi$, respectively. We implement the other easy gate Paulis with analogous Hamiltonians. We perform a circuit simulation by instantaneously switching between easy and hard rounds. The Hamiltonian $H_I$ also acts on the state with $J_k\sim \mathcal{N}(0, 10^{-3})$. We set the initial state to $\ket{\psi_0}=\ket{0_S}\ket{0_E}$ and calculate the error rate
\begin{equation}
	r(\noise)=1-\expval{\rho_\text{noisy}}{\psi_\text{ideal}} \,.
\end{equation}

In Fig.~\ref{fig:nonmarkov}, we plot the number of clock cycles ($M$) vs. the average error rate $r_\text{avg}=\avg\,r$. The data obtained for circuits $A$ and $B$ correspond to subfigures a) and b). In a), the bare average error rate grows quadratically while in b), the bare average error rate is quasilinear. After 100 cycles of gates, there is about a tenfold difference in the average error rate. We attribute the discrepancy to the observation that random hard gate rounds implement effective RDD \cite{Viola2005a} over a characteristic gate length and consequently suppress correlated errors. The RC data differs by less than 5\% between the two circuits, and in both cases, has a much lower error rate than the bare circuit.

\begin{figure}[t]
\includegraphics[width=\columnwidth]{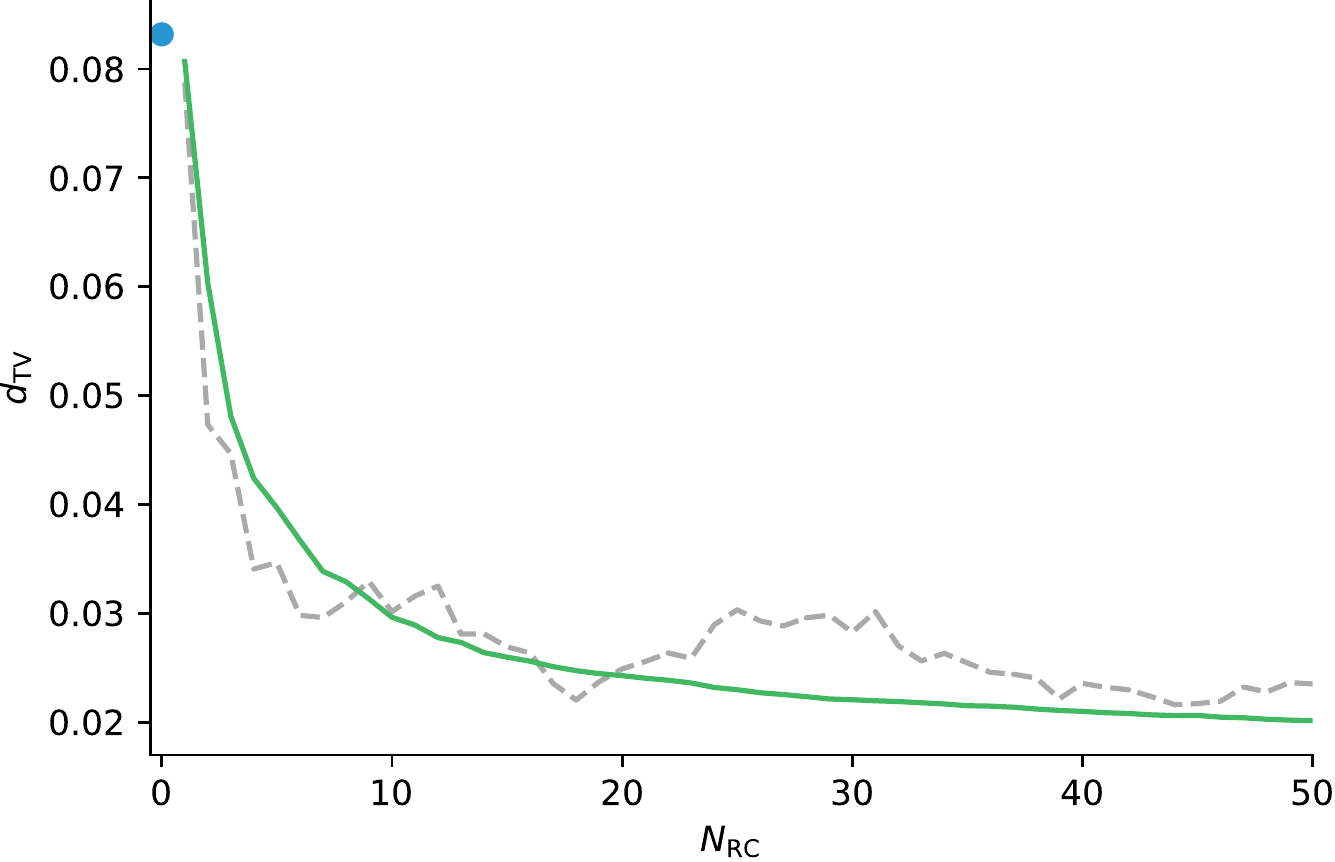}
\caption{Plots of the variational distance $d_\text{TV}$ for simulated data as a function of $N_\text{RC}$, the number of averaged randomly compiled circuits. Since we sample the gate set uniformly at random to construct the cycles in the raw circuit, there is effectively no difference in the average value of $d_\text{TV}$ for the raw circuit (blue circle) and the average for one random compilation (green line, $N_\text{RC}=1$). As we increase $N_\text{RC}$, $d_\text{TV}$ decreases significantly, indicating a reduction in the coherent component of the errors. The data points for the raw circuit and randomly compiled circuit are the averages of $10^2$ simulations. The grey dashed line indicates one such simulation.}
\label{fig:NumCompilations1}
\end{figure}

\begin{figure}[t]
\includegraphics[width=\columnwidth]{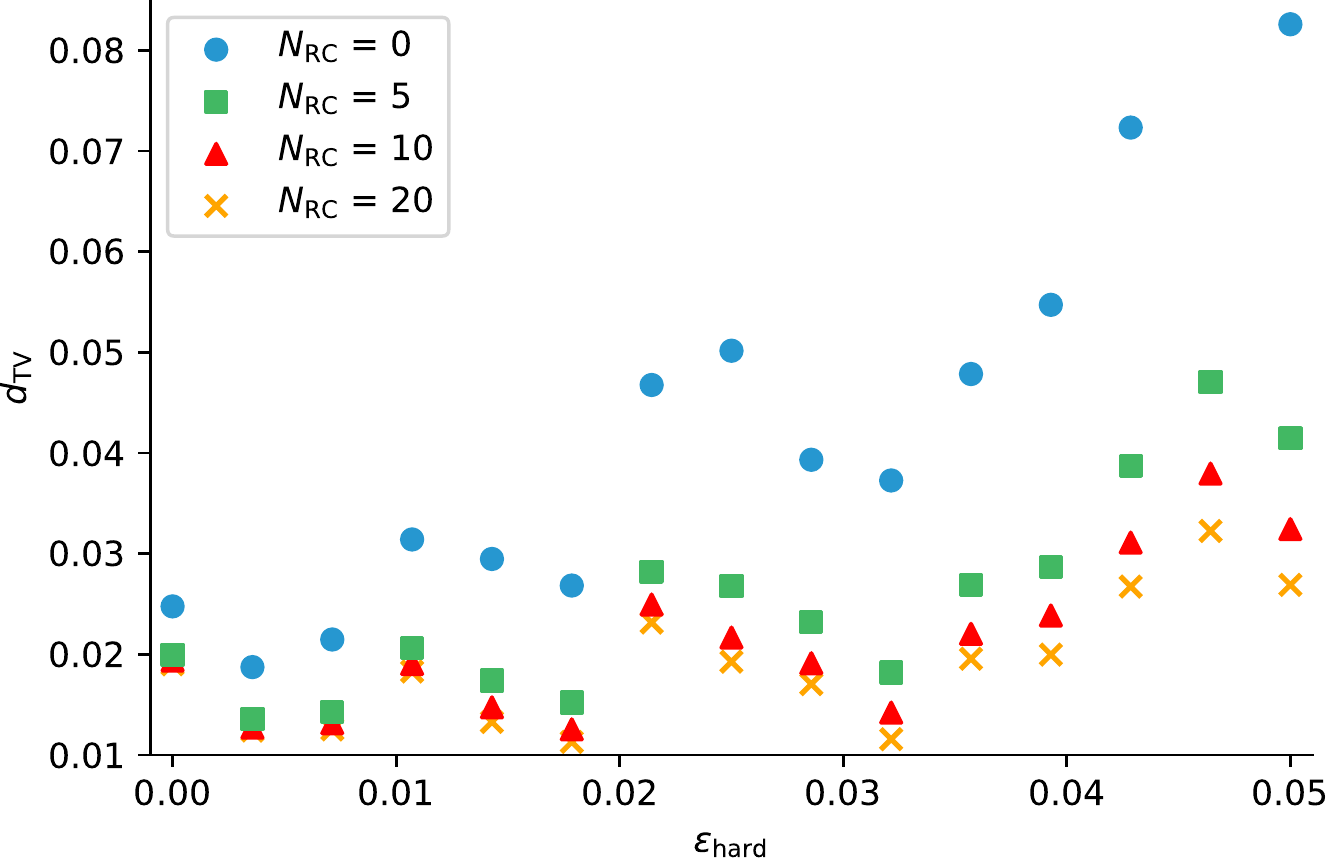}
\caption{Simulated data illustrating a progressive reduction in the variational distance $d_\text{TV}$ with increasing values of $N_\text{RC}$, the number of randomly compiled circuits that we average over. The relative improvement under RC is largest when $\epsilon_\text{hard}/\epsilon_\text{easy}$ is large. We plot $d_\text{TV}$ as a function of $\epsilon_\text{hard}$, the overrotation on the CNOT gates in the circuit. Easy gate rounds have gate-dependent coherent overrotations by $\epsilon_\text{easy} = 0.005$, which is highlighted by the gap between data sets that persists even when $\epsilon_\text{hard}=0$. We generate the incoherent part of the errors by modeling decoherence with somewhat pessimistic $T_1$ and $T_2$ times. The data points are the average of $10^2$ random circuit simulations.}
\label{fig:NumCompilations2}
\end{figure}

The above simulation highlights a distinct effect of RC that Ref.~\cite{Wallman2016a} does not report: a single randomly compiled circuit, on average, prevents the coherent accumulation of errors. If we studied uniformly random circuits when generating Fig.~\ref{fig:nonmarkov}, there would be no improvement under RC. The apparent invariance arises because $r$ is a linear functional and does not capture the error's off-diagonal components. However, there is another way that averaging over different randomized compilations helps: it ensures that results are stable. The error rate trajectory corresponding to one randomly compiled circuit may appear chaotic and differ from one compilation to the next. Averaging over several randomized compilations smooths out any such variations.

\begin{figure*}[t]
\includegraphics[width=\linewidth]{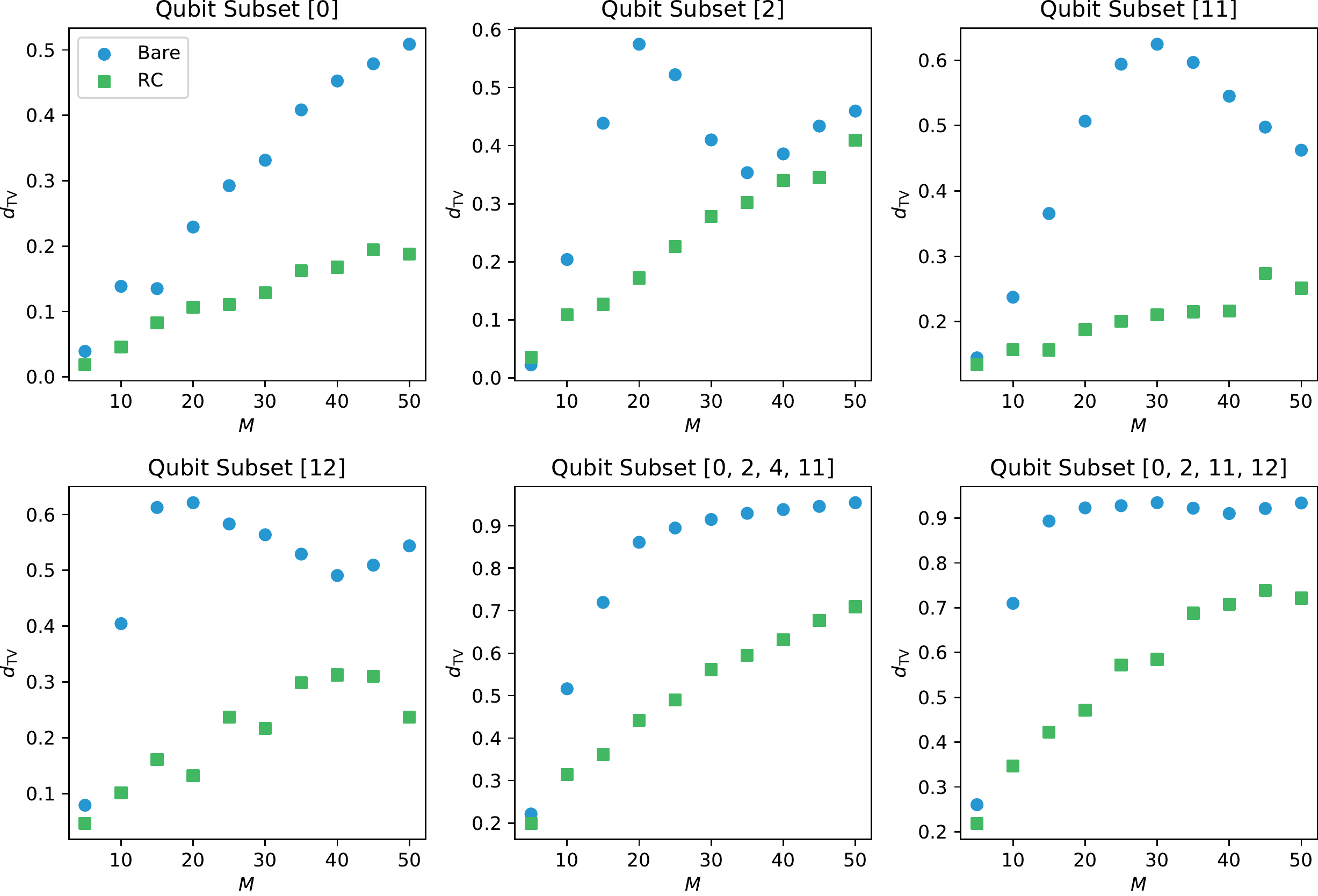}
\caption{Experimental data showing a clear improvement in the TVD under RC. We implement successive identity cycles on IBM's Melbourne chip sandwiched between two Hadamard gates and measure the TVD. Blue circles denote the bare circuit, and green squares mark the randomly compiled circuit. The title of each subplot denotes the measured qubits.}
\label{fig:RCHadamard}
\end{figure*}

\subsubsection{Noise tailoring property}

The second way in which RC improves errors is through its \emph{noise tailoring property}. When averaged over (approximately 5-10+) random circuit compilations, the errors acting on a particular circuit tends towards a stochastic Pauli channel.

It is informative to review RC's noise tailoring property under an elementary unitary error model. We consider an arbitrary single-qubit circuit that only induces an $X$ overrotation. Using the PTM representation of the overrotation error channel, the difference between the identity and the error channel in the ideal circuit frame is
\begin{equation}
	\Lambda - I =
	\begin{pmatrix}
		0 & 0 & 0 & 0 \\
		0 & 0 & 0 & 0 \\
		0 & 0 & 1-\cos\epsilon & \sin\epsilon \\
		0 & 0 & -\sin\epsilon & 1-\cos\epsilon
	\end{pmatrix} \,,
\end{equation}
where the unitary representation of the overrotation is $U=\exp(-i\epsilon X/2)$. In the limit $\epsilon \ll 1$, the magnitude of the off-diagonal elements is $\sin\epsilon\approx\epsilon$, while the diagonal elements are $1-\cos\epsilon\approx\epsilon^2/2$. Under RC, the off-diagonal elements of $\Lambda-I$ tend to zero as the number of random compilations approaches infinity. While completing this work, Ref.~\cite{Hashim2020a} was announced, and it presents a more detailed argument.

Linear error metrics, like fidelities, are strictly functions of the diagonal elements of the PTM representation of the effective error process, while nonlinear metrics typically depend nontrivially on the off-diagonal elements of the PTM description of the error. An easily measurable nonlinear quantity describing the errors affecting the implementation of a particular circuit is the total variational distance (TVD),
\begin{equation}
	d_\text{TV}(\mathcal{C}, \mathcal{C}_\text{ideal}) = \frac{1}{2}\sum_j\abs{\Pr(j|\mathcal{C})-\Pr(j|\mathcal{C}_\text{ideal})} \,,
\end{equation}
where $\mathcal{C}_\text{ideal}$ and $\mathcal{C}$ are the ideal and actual circuits. The quantity measures the distance between the noisy and ideal probability distributions with respect to the computational basis.

The nonlinearity of the TVD is related to the uniformity of the ideal probability distribution. If the distribution comprises a single computational basis state, then the TVD is linear and independent of the off-diagonal components of the PTM. Meanwhile, when the ideal distribution is close to the uniform distribution, the TVD is highly nonlinear and depends significantly on off-diagonal elements of the PTM. Ref.~\cite{Hashim2020a} experimentally studies the dependence in more detail.

\begin{figure*}[t]
\includegraphics[width=\linewidth]{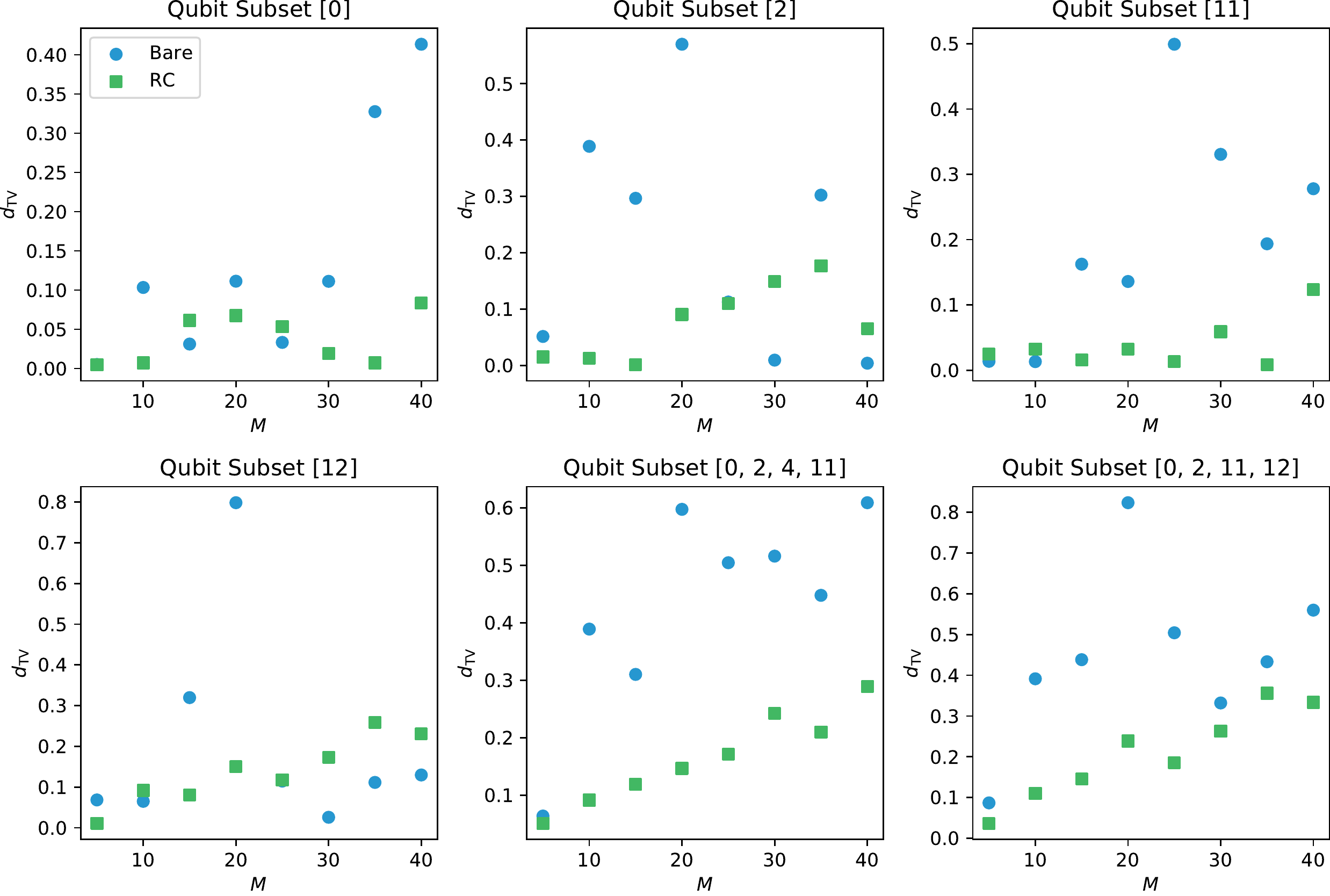}
\caption{TVD data illustrating an improvement arising from the noise tailoring property of randomized compiling. We implement rounds of simultaneous random elements of $\su{2}$ on all qubits and estimate the TVD. Blue circles denote the bare circuit, and green squares mark the randomly compiled circuit. The measured qubits in each subplot are specifically those considered in the preceding figure.}
\label{fig:RCSU2}
\end{figure*}

Consider the regime where the TVD is nonlinear. In the small $\epsilon$ approximation, the scaling of $d_\text{TV}$ improves from $\epsilon$ and $\epsilon^2$. In terms of the TVD, the scaling $d_\text{TV} \sim \sqrt{r}$ is upgraded to $d_\text{TV} \sim r$. In the next decade, processors will likely have error rates on the order of $10^{-3}$. In the case that these errors are mostly coherent, RC can yield $d_\text{TV} \sim 10^{-6}$. In our experiments with public IBM chips, there are massive errors, and the improvement emerging from the noise tailoring property of RC is limited. Despite the large errors, our experiments show a definite positive effect.

\subsubsection{Example: Uniformly random cirucits}

In this example, we model $N=6$ qubits on a digital computer and initialize the system in the $\ket{0}^{\otimes N}$ state. We introduce coherent and incoherent errors comparable to those seen in state-of-the-art superconducting processors. The qubits in our system experience elementary decoherence with $T_1=50\,\mu\text{s}$ and $T_2=50\,\mu\text{s}$. The duration of a single-qubit gate is $t_\text{single}=25\,\text{ns}$, while a multi-qubit gate takes $t_\text{multi}=100\,\text{ns}$. There are significant gate-dependent errors. Rather than implementing a gate $U$, we apply the overrotated operation $U_\text{err}=U^{1+\epsilon}$. We investigate two families of uniformly random circuits.

In our first simulation, we introduce overrotation errors with single-qubit overrotations by $\epsilon_\text{easy} = 0.01$ and two-qubit overrotations by $\epsilon_\text{hard}=0.05$. We implement circuits with $M=5$ rounds. Easy gate cycles implement random elements of $\su{2}^{\otimes N}$, while hard gate cycles sample uniformly at random from the set of hard gates generated by the $\Delta(X)$ gate. We compute the TVD between the ideal and noisy outputs under $N_\text{RC}$ randomized compilations. To get smooth TVD decays, we repeat this simulation $10^2$ times.

We plot the resulting TVD values in Fig.~\ref{fig:NumCompilations1}. Because the target circuits are random, the dynamical decoupling property does not improve the result of this simulation, and the value of $d_\text{TV}$ is essentially the same for $N_\text{RC}=0$ and $N_\text{RC}=1$. As we increase $N_\text{RC}$, noise tailoring manifests and $d_\text{TV}$ drops. Most of the reduction appears by $N_\text{RC}=10$, with a marginal improvement at and beyond $N_\text{RC} = 20$.

In our second numerical experiment, we fix $\epsilon_\text{easy}=0.005$ and vary $\epsilon_\text{hard}$. We sample the same set of circuits described in the experiment above and perform the same number of simulations. For a fixed value of $N_\text{RC}$ we estimate $d_\text{TV}$ values for different ratios of $\epsilon_\text{hard}$/$\epsilon_\text{easy}$ and plot the results in Fig.~\ref{fig:NumCompilations2}. When $\epsilon_\text{hard}$/$\epsilon_\text{easy}$ is small, RC has a noticeable but small effect on the TVD since it only tailors the gate-independent part of the easy gate noise. Contrarily, when $\epsilon_\text{hard}$/$\epsilon_\text{easy}$ is large, RC greatly reduces the TVD since the gate-dependent part of the easy gate noise is relatively small.

\section{Experimental Data} \label{sec:Experiments}

In this section, we describe data obtained using the IBM Quantum Experience platform that proves the real experimental benefits of RC. These observations extend earlier experimental results practicing Pauli frame randomization \cite{Ware2018}. Note that as experimental error rates decrease, the relative improvement under RC will massively increase, as implied by the preceding discussion.

\subsection{Single-qubit gates}

The first IBM chip that we use is a 14 qubit device, Melbourne. From preliminary diagnostics, we identify the presence of massive errors. In order to see an improvement under RC, the error rate must be sufficiently low. That is, the error rate $r$ should be significantly smaller than 1.

\subsubsection{Parallel trivial single-qubit gates}

Because of the large errors, we naturally start by studying a simple family of trivial circuits. Hard cycles implement physical identity gates while easy cycles implement simultaneous Hadamard gates on all 14 qubits. The easy cycles cause considerable coherent crosstalk. We construct circuits with various numbers of (identical) clock cycles ($M$), and randomly compile each bare circuit $N_\text{RC}=18$ times. We marginalize over six different qubit subsets to highlight diverse local TVD behavior and plot the results in Fig.~\ref{fig:RCHadamard}.

In all subplots, we observe two key ways that RC improves device performance:
\begin{enumerate}
\item The TVD as a function of the circuit depth is approximately linear under RC when it is not too large. In contrast, the raw TVD is highly nonlinear, and the performance of a circuit is not easily predictable.
\item For every pair of plotted data points, the TVD under RC is lower than the bare TVD.
\end{enumerate}

\subsubsection{Parallel random single-qubit gates}

Next, we investigate a more complicated set of circuits that are structurally similar to those described in the preceding experiment. Hard gates again consist of physical identity gates, but each easy gate cycle implements random elements of $\su{2}^{\otimes 14}$. We marginalize over the same qubit subsets as above and plot the results in Fig.~\ref{fig:RCSU2}.

Compared to the raw circuit data in Fig.\ref{fig:RCHadamard}, the raw data in Fig.~\ref{fig:RCSU2} is much more chaotic. The unpredictable behavior arises because a random sequence of gates from $\su{2}$ natively performs RDD. Since the circuit samples a single random dynamical decoupling sequence, the control follows a random walk \cite{Viola2005a}. The average under many random walks is quasi-monotonic, yet a single instance is often turbulent and explains our data. Under RC, the TVD improves similarly to above. The resulting TVD points are close to one another, and there is a sharp average decline in the TVD.

%

\subsection{Entangling gates}

\subsubsection{Parallel entangling gates}

Having tested RC using physical parallel single-qubit gates, we now examine its effect on real simultaneous CNOT gates. Easy gate rounds in our circuit implement random elements of $\su{2}^{\otimes 14}$ like in the preceding experiment, but the hard gate rounds are no longer trivial. Every hard cycle applies CNOT operations to three pairs of qubits: $(0,1)$, $(2,3)$ and $(4,5)$. The bare circuit already contains random single-qubit gates, so the noise tailoring property of RC is the only way that it can improve the device output.

\begin{figure}[t]
\includegraphics[width=\linewidth]{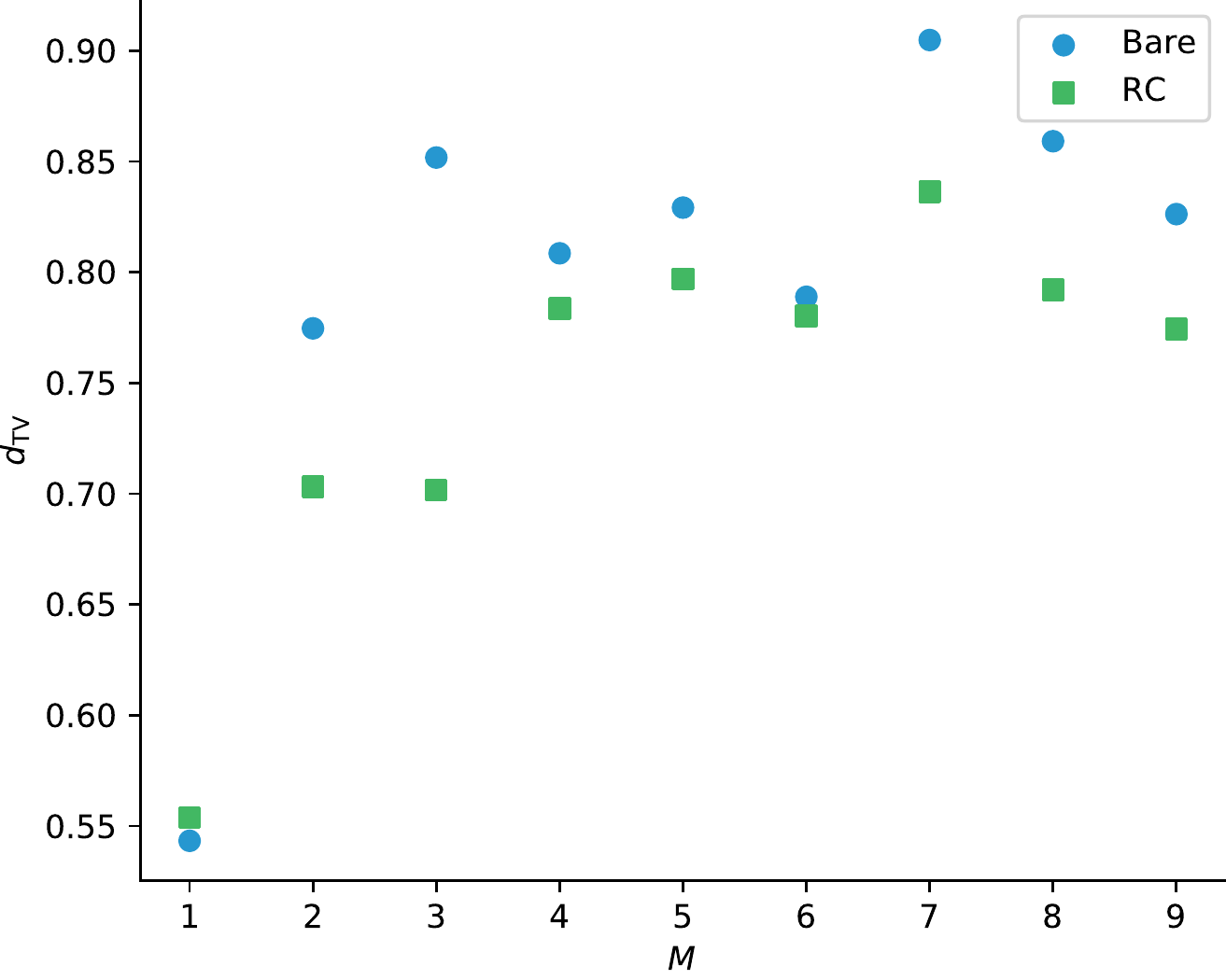}
\caption{Data showing reduced TVD values following the application of RC to a circuit where hard rounds perform simultaneous entangling gates. Each hard rounds applies CNOT operations to three pairs of qubits: $(0,1)$, $(2,3)$ and $(4,5)$. Between the parallel CNOT cycles, we run random single $\su{2}$ operations on all 14 qubits. Blue circles denote the bare circuit, and green squares mark the randomly compiled circuits.}
\label{fig:RCCNOT}
\end{figure}

We plot the TVD data corresponding to the three pairs of qubits in Fig.~\ref{fig:RCCNOT}. Despite the large errors that push our system outside the regime where improvements are theoretically guaranteed, we see a 5-10\% reduction for most circuit lengths. This data suggests that RC provides a performance boost to both very noisy devices and excessively deep circuits.

\subsubsection{Quantum chemistry experiment}

Quantum chemistry is among the most promising potential applications of large-scale fault-tolerant quantum computers. We examine the result of applying IBM Melbourne to an electronic structure problem involving the molecule lithium hydride (LiH) and generate quantum circuits using the methods of Ref. \cite{Ryabinkin2018,Ryabinkin2020}. Compared to a simpler molecule like $\text{H}_2$,  LiH has significantly more complex orbitals.

These circuits utilize the qubit coupled-cluster (QCC) method within the variational quantum eigensolver (VQE) formalism. QCC produces compact, economical circuits that model highly entangled systems on hardware limited by the entanglement of only (noisy) two-qubit gates. In our LiH experiment, for example, we need only employ six 2-qubit CNOT gates. We apply RC with $N_\text{RC}=36$ randomized compilations.

In contrast to the circuits in the two prior experiments, the QCC circuits are structured, suggesting that both the decoupling and noise tailoring aspects of randomized compiling will enhance the measured statistics. Unlike in the simultaneous CNOT experiment, there is an improvement for all independent values. RC data is better by an average of more than 25\%.

\begin{figure*}[t]
\includegraphics[width=\linewidth]{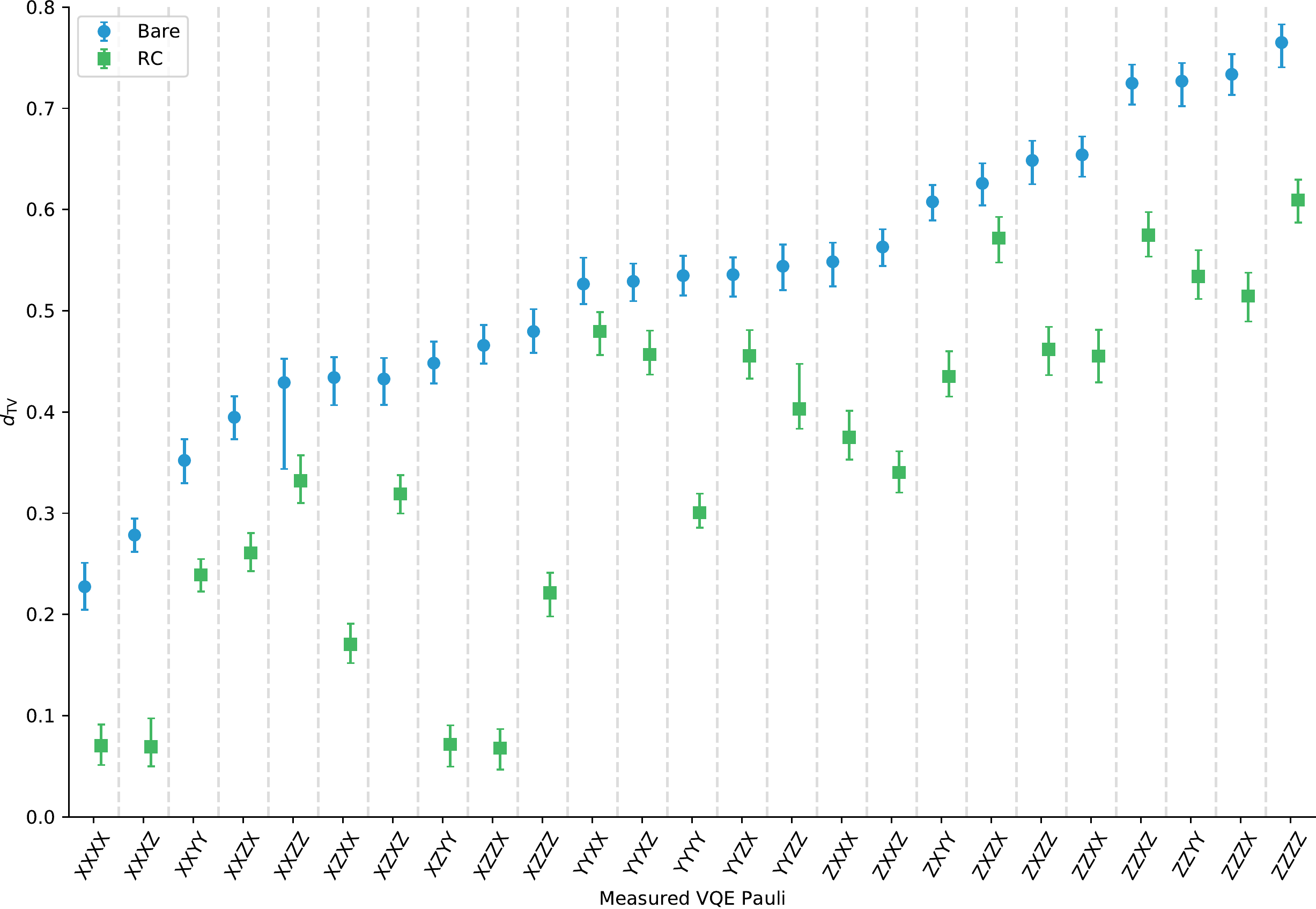}
\caption{Experimental TVD estimates corresponding to an electronic structure problem involving the molecule LiH. Blue circles mark the raw data, while green squares mark the randomly compiled data. The error bars specify 95\% confidence intervals. For every measured Pauli word (tensor products of Pauli matrices), RC improves the TVD, with an average improvement of more than 25\%.}
\label{fig:RCCNOTLadder}
\end{figure*}

\section{Conclusion} \label{sec:Conclusion}

In this paper, we have extended the theory of RC and shown its capacity to improve real experimental devices. On the theory side, our results show that RC is a robust tool for passively reducing the effects of both non-Markovian and time-dependent errors. We also detailed a close connection between RC and random dynamical decoupling, and incorporating hybrid-deterministic techniques analogous to hybrid dynamical decoupling \cite{Santos2006,Khodjasteh2008} may be a natural extension for RC. On the experimental side, we applied RC to a diverse set of circuits and found unequivocally better TVD estimates. Not only does this data affirm the value of RC, but it also shows that RC can improve noisy circuits in the high-error regime beyond where our theoretical results guarantee improvements.

\section{Acknowledgments}

We thank OTI Lumionics Inc. for providing VQE circuits and IBM for providing access to IBM Quantum Experience. This research was supported by the U.S. Army Research Office through grant W911NF-21-1-0007, Transformative Quantum Technologies, and Quantum Benchmark Inc.

\bibliographystyle{apsrev4-1}
\input{references.bbl}

\end{document}

%% file: references.bbl
%